\documentclass[aps,twocolumn,showpacs,showkeys,amsmath,amssymb,prb]{revtex4} 

\usepackage{graphicx}   
\usepackage{dcolumn}    
\usepackage{bm}         
\usepackage{natbib}      
\usepackage{float}

\begin{document} 

\title{Process chain approach to the Bose-Hubbard model: 
	Ground-state properties and phase diagram}

\author{Niklas Teichmann} \email{teichmann@theorie.physik.uni-oldenburg.de} 
\author{Dennis Hinrichs}
\author{Martin Holthaus}
\affiliation{Institut f\"ur Physik, Carl von Ossietzky Universit\"at, 
	D-26111 Oldenburg, Germany}
\author{Andr\'e Eckardt}
\affiliation{ICFO -- Institut de Ci\`{e}ncies Fot\`{o}niques, 
	Mediterranean Technology Park, 
	E-08860 Castelldefels (Barcelona), Spain}
\date{April 1, 2009}

\begin{abstract}
	We carry out a perturbative analysis, of high order in the tunneling
	parameter, of the ground state of the homogeneous Bose-Hubbard model 
	in the Mott insulator phase. This is made possible by a diagrammatic 
	process chain approach, derived from Kato's representation of the 
	many-body perturbation series, which can be implemented numerically 
	in a straightforward manner. We compute ground-state energies, 
	atom-atom correlation functions, density-density correlations, and 
	occupation number fluctuations, for one-, two-, and three-dimensional 
	lattices with arbitrary integer filling. A phenomenological scaling 
	behavior is found which renders the data almost independent of the 
	filling factor. In addition, the process chain approach is employed 
	for calculating the boundary between the Mott insulator phase and the 
	superfluid phase with high accuracy. We also consider systems with 
	dimensionalities $d > 3$, thus monitoring the approach to the 
	mean-field limit. The versatility of the method suggests further 
	applications to other systems which are less well understood.
\end{abstract}



\maketitle

\section{Introduction} 

With the seminal experiment by Greiner~\emph{et al.},~\cite{GreinerEtAl02} 
who have observed the quantum phase transition from a Mott insulator to a 
superfluid~\cite{Sachdev99} in a gas of ultracold Rubidium atoms stored in a 
three-dimensional optical lattice, the regime of strongly interacting Bose 
gases has become accessible. Due to its perfectly controllable parameters, the 
experimental set-up, as suggested by Jaksch~\emph{et al.},~\cite{JakschEtAl98} 
provides a fantastic testing ground for quantum many-body 
physics.~\cite{BlochEtAl08} Meanwhile, the transition has also been observed 
in one- and two-dimensional optical lattices.~\cite{StoeferleEtAl2004,
SpielmanEtAl2007} Ultracold atoms in optical lattices are described by the 
Bose-Hubbard model,~\cite{FisherEtAl89} which incorporates two competing trends
in an elementary manner: While the repulsive interaction between the atoms 
tends to localize the particles at individual sites of the lattice potential, 
tunneling between neighboring sites favors delocalization, and tends to 
suppress phase fluctuations.

The Bose-Hubbard model and its descendants have been intensively studied 
within the last years. Important techniques for monitoring its ground-state 
properties and the phase diagram include the mean-field 
approach,~\cite{FisherEtAl89} strong-coupling 
expansions,~\cite{FreericksMonien96,ElstnerMonien99}
methods using the density matrix renormalization 
group~(DMRG),~\cite{KuehnerEtAl1998,RapschEtAl1999,KollathEtAl04,KollathEtAl07}
and, more recently, quantum Monte Carlo (QMC) 
simulations.~\cite{BatrouniEtAl1992,WesselEtAl2004,CapogrossoSansone07,
CapogrossoSansone08}

In the present work we apply a recently suggested process chain 
approach~\cite{Eckardt08} to the $d$-dimensional homogeneous Bose-Hubbard 
model, in order to investigate the properties of its ground state in the Mott 
insulator phase for any integer filling factor $g \ge 1$, as well as the phase 
diagram for $d\ge 2$, by means of a high-order expansion in the tunneling 
parameter. To achieve this, we have turned Kato's representation of the 
perturbation series~\cite{Kato49} into a numerically executable algorithm
which handles symbolic diagrams as inputs. Order by order, each observable 
then is represented by a set of such diagrams, equipped with appropriate 
weight factors.

The paper is organized as follows: In Sec.~\ref{sec:model} the Bose-Hubbard 
model is briefly recapitulated. Our zeroth-order Hamiltonian contains the 
local on-site interaction only, while the tunneling term will be treated 
perturbatively. Section~\ref{sec:Kato} introduces Kato's formulation 
of the perturbation series, and explains the required elements of the 
diagrammatic process chain approach.~\cite{Eckardt08} As an instructive
example for this technique, we outline in some detail how to calculate the 
fourth-order energy correction induced by the tunneling term. We also discuss 
how fully correlated ground-state expectation values are accessible via the 
perturbational approach. In Sec.~\ref{sec:results} we present results for the 
ground-state energy, correlation functions, and occupation number fluctuations,
as obtained for homogeneous hypercubic lattices with dimensionality $d=1$, 
$2$, and $3$. The required diagrams are developed, and a phenomenological 
scaling is suggested, which makes the data almost independent of the filling 
factor~$g$. Results typically are calculated up to tenth order in the tunneling
parameter. For $d=1$ and unit filling we obtain perfect agreement with the  
results of the high-order symbolic perturbative expansion of Damski and 
Zakrzewski.~\cite{DamskiEtAl2006} We follow their instructive work to some
extent in spirit, opening up the regimes of higher~$d$ for any~$g$. 
The Mott insulator-to-superfluid phase transition is then discussed in 
Sec.~\ref{sec:phase} within the framework of the process chain approach.
The phase boundary is determined by invoking the method of the effective
potential which provides a convenient indicator for the transition 
point;~\cite{NegeleOrland98,SantosPelster08} this leads to a computational 
scheme which again can be expressed in terms of  
diagrams.~\cite{TeichmannEtAl2009} That scheme is worked out exactly in the 
case of infinite lattice dimensionality~$d$, and then reproduces the well-known
mean-field result. For $d < \infty$ our data, obtained by numerically 
evaluating the diagrams up to some order~$\nu$, lend themselves to a simple 
extrapolation procedure for $\nu \to \infty$. Critical parameters are 
computed in this manner for $d = 2$ and $d = 3$ and any filling factor~$g$; 
for $g = 1$, they compare quite favorably to recent high-precision QMC
data.~\cite{CapogrossoSansone08,CapogrossoSansone07} 
The last part of this Sec.~\ref{sec:phase} details how the mean-field 
prediction is approached with increasing dimensionality of the system.
We close the paper with some concluding remarks and a short outlook in 
Sec~\ref{sec:conclusion}.

\section{The model} 
\label{sec:model} 

Ultracold atoms in optical lattices are fairly well described by a single-band 
Bose-Hubbard model. \emph{Ultracold} in this context means that the thermal 
de Broglie wavelength is at least equal to the lattice constant, i.e., to 
half the wavelength of the laser radiation generating the lattice. The 
assumptions underlying this model, requiring that the thermal and the 
interaction energies be smaller than the gap between the lowest and the first 
excited Bloch band, are fulfilled if the lattice is sufficiently deep. 
Denoting the interaction energy of a pair of particles occupying the same 
lattice site by $U$, the chemical potential at site $i$ by $\mu_i$, and the 
hopping matrix element connecting well~$i$ and well~$j$ by $J_{ij}$, the model 
takes the form
\begin{equation}
	H = \frac{U}{2} \sum_{i} \hat{n}_i(\hat{n}_i-1)  
	- \sum_{i,j} J_{ij}\, \hat{a}_i^{\dagger} \hat{a}_j^{\phantom \dagger}
	-  \sum_{i}\mu_i \hat{n}_i \;.
\end{equation} 
Here $\hat{a}_i^{\dagger}$ and $\hat{a}_i^{\phantom \dagger}$ are the 
bosonic creation and annihilation operators for site $i$, and 
$\hat{n}_i = \hat{a}_i^{\dagger} \hat{a}_i^{\phantom \dagger}$ is the 
corresponding number operator. The chemical potential $\mu_i$ can incorporate 
an arbitrary confining potential, and then depends on the lattice site. By 
choosing appropriate hopping elements $J_{ij}$, longer-range or anisotropic 
hopping can be modeled. In this study we stick to the paradigmatically 
simple case with site-independent chemical potential $\mu$ and isotropic 
nearest-neighbor hopping of positive strength~$J$ on a hypercubic lattice. 
Utilizing the interaction energy~$U$ as the energy scale of reference, 
the dimensionless Hamiltonian then reads as
\begin{equation}
	H_{\rm BH} = H_0 + H_{\rm tun} \; ,
\label{eq:Hamiltonian}
\end{equation} 
where 
\begin{equation}
	H_0 = \frac{1}{2} \sum_{i \phantom \rangle} 
	\hat{n}_i(\hat{n}_i-1) - \mu/U \sum_i \hat{n}_i	
\end{equation} 
is site-diagonal, and
\begin{equation}
	H_{\rm tun} = - J/U \sum_{\langle i,j \rangle} 
	\hat{a}_i^{\dagger} \hat{a}_j^{\phantom \dagger}
\label{eq:Htun}
\end{equation} 
describes tunneling between adjacent sites, with $\langle i,j \rangle$ 
indicating that the sum is restricted to nearest neighbors. One can 
easily oversee two limiting cases: If the lattice is very deep, tunneling 
even between neighboring wells is inhibited, because the tunneling 
parameter~$J/U$ vanishes exponentially with increasing lattice 
depth.~\cite{Zwerger2003,BoersEtAl2007} The sites then decouple and the 
Hamiltonian becomes local; one only needs to consider $H_0$. Minimizing 
the on-site energy $\varepsilon_i = n_i(n_i-1)/2 -  n_i\mu/U$ of a system 
with $J/U = 0$ leads to the site-occupancies~\cite{FisherEtAl89} 
\begin{equation}
	n_i = \left\{\begin{array}{llll}
	0 & \text{for } & \mu/U < 0 & \\            
	g & \text{for } & g-1<\mu/U<g \; , \; & 1\leq g \in \mathbb N \; .	
	    \end{array} \right. 
\end{equation} 
Thus, as long as the chemical potential $\mu/U$ falls between $g-1$ and $g$, 
each site is occupied by an integer number $g = N/M$ of atoms, where $N$ is 
the total number of particles, and $M$ the number of lattice sites. Denoting 
the vacuum state by $|0\rangle$, the $H_0$ ground state then is given by the 
product state
\begin{equation}
	| \mathbf m \rangle = \prod_{i=1}^M 
	\frac{\left(\hat{a}_i^{\dagger}\right)^g}{\sqrt{g!}} |0 \rangle \; . 
\label{eq:GroundStateMI}
\end{equation} 
For integer $\mu/U$ the ground state is $2^M$-fold degenerate. The parameter 
regime $g J/U \ll 1$ gives rise to insulating phases, because the system 
remains incompressible for small, but non-zero tunneling strength. A small 
change of the chemical potential~$\mu$ then does not lead to a change of the 
occupation number: $\partial \langle \hat{n}_i \rangle/\partial \mu = 0$.

The opposite limiting case appears when the interaction between the particles 
can be neglected in comparison with the kinetic energy, $g J/U \gg 1$. The 
ground state then becomes an ideal Bose-Einstein condensate with all particles 
occupying the zero-quasimomentum state of the lowest band:
\begin{equation}
	|  \mathbf{sf} \rangle = \frac{1}{\sqrt{N!}}
	\left(\frac{1}{\sqrt{M}}\sum_{i=1}^M 
	\hat{a}_i^{\dagger} \right)^N|0 \rangle 
	\;. 
\label{eq:GroundStateSF}
\end{equation} 
Observe that this is an eigenstate of $H_{\rm tun}$. Nonetheless, 
in what follows we use the site-diagonal Hamiltonian~$H_0$ and the Fock 
state~(\ref{eq:GroundStateMI}) as the starting point for our perturbative 
analysis.

\section{Kato formalism and process chain approach} 
\label{sec:Kato} 

For calculating corrections to the ground-state energy, and further 
ground-state expectation values, we resort to the representation of the 
perturbation series given by Kato.~\cite{Kato49,MessiahII} Its distinct 
advantage lies in the fact that it yields closed expressions for the 
perturbative corrections in any order, in contrast to the more familiar
recursive formulation of the Rayleigh-Schr\"odinger perturbation 
series.~\cite{Baym69,GelfandEtAl90}

The ground state $|\mathbf{m}\rangle$ of the Hamiltonian~$H_0$ is a 
product of local Fock states with $g$ particles sitting at each site. 
When this system is subjected to some perturbation~$V$, not necessarily 
given by $H_{\rm tun}$, the $n$th-order correction to its energy 
is given by the trace~\cite{Kato49,MessiahII}
\begin{eqnarray}	
	E_{\mathbf{m}}^{(n)} = {\rm tr} \left[ \sum_{ \{\alpha_\ell\} } 
	S^{\alpha_1} V S^{\alpha_2} V S^{\alpha_3} \ldots 
	S^{\alpha_n}VS^{\alpha_{n+1}} \right] \; ,
\label{eq:Kato_energy}
\end{eqnarray}
where the sum runs over all possible sequences $\{\alpha_\ell\}$ of
$n+1$ non-negative integers which obey the condition 
\begin{equation}
	\sum_{\ell=1}^{n+1} \alpha_\ell = n-1 \; .	
\label{eq:Kato_condition}
\end{equation}
The operators $S^{\alpha}$ linking the individual perturbation events~$V$ 
are given by
\begin{equation}
S^{\alpha} = \left\{\begin{matrix} 
	-|\mathbf{m}\rangle \langle \mathbf{m}| & 
	\quad \text{for } \alpha = 0 \\
	\displaystyle
	\sum\limits_{i\neq \mathbf{m}} 
	\frac{|i \rangle \langle i|}
	     {(E_{\mathbf{m}}^{(0)} - E_i^{(0)})^{\alpha}} & 
	\quad \text{for } \alpha > 0            
	\end{matrix}\right.  
\label{eq:Kato_S}
\end{equation}
with the energies $E_{\mathbf{m}}^{(0)} = M\left[g(g-1)/2-(\mu/U) g \right]$ 
and $E_i^{(0)} = \sum_i \left[ n_i(n_i-1)/2 - (\mu/U) n_i  \right] $ of the 
unperturbed $H_0$ eigenstates $|\mathbf{m}\rangle$ and $|i\rangle$, 
respectively. Because these Fock states form an orthonormal basis, one easily 
derives
\begin{eqnarray}	
	\begin{array}{llll}			
	S^0 S^{\alpha} &       =  0                
	& \text{for } & \alpha > 0 \\
 	S^{\alpha} S^{\beta} & = S^{\alpha+\beta}  
	& \text{for } & \alpha, \beta > 0 \; .
	\end{array}
 \label{eq:S_rel}	
\end{eqnarray}	
To see how Eq.~(\ref{eq:Kato_energy}) works, let us consider the 
energy correction in second order, i.e., for $n = 2$. The partition 
problem~(\ref{eq:Kato_condition}) then has the solutions 
$\{1,0,0\}$, $\{0,1,0\}$ and $\{0,0,1\}$. Accordingly, one finds
\begin{eqnarray}
	E_{\mathbf{m}}^{(2)} & = & {\rm tr} 
	\left[S^1 V S^0 V S^0 + S^0 V S^1 V S^0 + S^0 V S^0 V S^1 \right] 
	\nonumber \\
	& = & {\rm tr} \left[S^0 V S^1 V S^0\right] 
	\nonumber \\
	& = & \langle \mathbf{m}|   V S^1 V |\mathbf{m} \rangle 
	\nonumber \\	
	&=& \sum\limits_{i\neq \mathbf{m}}	
	\frac{\langle \mathbf{m}| V |i \rangle \langle i| V |\mathbf{m} \rangle}
	     {E_{\mathbf{m}}^{(0)} - E_i^{(0)}} \; .
\label{eq:E_2nd}	
\end{eqnarray} 
In the second step, cyclic interchangeability of operators under a trace 
has been used, together with Eq.~(\ref{eq:S_rel}).  The final 
expression~(\ref{eq:E_2nd}) coincides exactly with the familiar textbook 
result provided by the Rayleigh-Schr\"odinger perturbation theory (see, 
e.g., Refs.~\onlinecite{MessiahII,Baym69}), as it should. 

Due to the restriction~(\ref{eq:Kato_condition}), in any order~$n$ at 
least two superscripts~$\alpha_\ell$ are equal to zero, so that the 
trace~(\ref{eq:Kato_energy}) can always be rewritten as a sum of matrix 
elements of the standard form $\langle \mathbf{m} | V S^{\alpha_1} V \ldots 
V  S^{\alpha_{n-1}} V | \mathbf{m} \rangle$. Such elements will be 
called \emph{Kato terms}. We regard each Kato term as a sum over 
\emph{process chains}~\cite{Eckardt08} leading from $|\mathbf{m} \rangle$ 
back to $|\mathbf{m} \rangle$, with the individual processes corresponding 
to non-zero matrix elements $\langle i |V| j \rangle$. 

\begin{figure}
\includegraphics[scale=0.5]{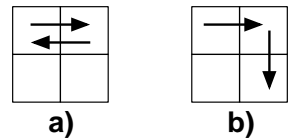}
\caption{Second-order tunneling processes on a lattice. While path~a) 
	contributes, path~b) gives no contribution to the energy correction, 
	as final and initial state do not coincide.} 
\label{fig:diagram_e_simple}
\end{figure}

In particular, when calculating energy corrections we identify each process 
with a term of $H_{\rm tun}$, setting
\begin{equation}
	V = - J/U \sum_{\langle i,j \rangle} 
	\hat{a}_i^{\dagger} \hat{a}_j^{\phantom \dagger}
	\;.
\label{eq:V_tun}
\end{equation} 
Each Kato term now can be viewed as a sum over certain chains of tunneling 
processes on the lattice. Because these Kato terms represent expectation 
values with respect to the state $|\mathbf{m} \rangle$, each chain 
has to start and to end in the state $|\mathbf{m} \rangle$. Thus, only closed 
loops of tunneling processes contribute to the energy correction. A simple 
example may illustrate this fact: Consider the energy correction in second 
order, given by Eq.~(\ref{eq:E_2nd}). One then has two tunneling processes, 
which could take place anywhere on the lattice. But only process chains 
for which initial and final state both coincide with $|\mathbf{m} \rangle$ 
give a contribution. This requires to tunnel back and forth, thus producing
the only closed loop with two individual tunneling processes, as illustrated
in Fig.~\ref{fig:diagram_e_simple}. Such closed loops of tunneling processes 
will be denoted as \emph{paths} in the following.
The respective \emph{sequence} of the individual processes, i.e., their
ordering, is quite important for the evaluation of the matrix elements, 
as will become evident soon. 

\begin{figure} 
\includegraphics[scale=0.5]{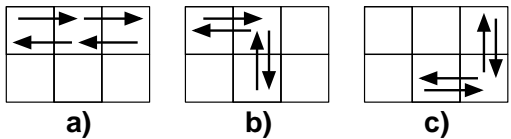}
\caption{Examples of topologically identical paths occurring in fourth order 
	when calculating the energy correction. These paths are described
	by the same diagram.}
\label{fig:paths_topological_identical}
\end{figure}

The particular perturbation~(\ref{eq:V_tun}) gives no contributions in odd 
orders, because no closed loops can be formed with an odd number of tunneling 
processes on a cubic lattice. In fourth order, the general Kato terms are
\begin{eqnarray}
	E_{\mathbf{m}}^{(4)} = &   &
	\langle \mathbf{m}|V S^0 V S^3 V S^0 V |\mathbf{m} \rangle
	\nonumber\\
	+ & 2&\langle \mathbf{m}|V S^2 V S^1 V S^0 V |\mathbf{m} \rangle 
	\nonumber\\
 	+ & &\langle \mathbf{m}|V S^2 V S^0 V S^1 V |\mathbf{m} \rangle	 
	\nonumber\\
        + & &\langle \mathbf{m}|V S^1 V S^1 V S^1 V |\mathbf{m} \rangle
	\; .		
\label{eq:E_4th}
\end{eqnarray} 
The first term requires that the initial state $|\mathbf{m} \rangle$ 
be recovered after the first process (numbered from right to left), 
because  $S^0 = -|\mathbf{m} \rangle \langle \mathbf{m} |$ occurs. 
With the perturbation~(\ref{eq:V_tun}) this is impossible. Hence, when 
treating perturbations that vanish in first order, like the tunneling
events~(\ref{eq:V_tun}), one can further reduce the number of Kato terms.
To the third term in Eq.~(\ref{eq:E_4th}) only chains revisiting 
$|\mathbf m \rangle$ after two processes contribute, while evaluating 
the fourth term requires to take into account all those permutations 
of the processes forming the closed loop which do not feature the state 
$|\mathbf{m} \rangle$ as an intermediate state. With four tunneling 
processes one can form lots of closed loops on the lattice, but as the 
system is homogeneous, paths which are topologically identical contribute 
in the same way. We subsume those topologically identical paths under a 
\emph{diagram}. Examples of topologically identical paths are sketched in
Fig.~\ref{fig:paths_topological_identical}. According to the linked cluster 
theorem,~\cite{GelfandEtAl90} disconnected diagrams do not contribute.
With every diagram we associate a \emph{weight factor} which incorporates, 
on the one hand, the subsummation of topologically identical paths and, 
on the other, an additional factor of $s^{-1}$ for a diagram visiting $s$ 
lattice sites, as each of these $s$~sites can be the ``origin'' of the diagram.

Thus, when calculating energy corrections the number~$\nu$ of tunneling
processes agrees with the respective order~$n$ of the perturbation series,
and the overall program for determining these corrections on a hypercubic 
lattice to a given order~$\nu$ consists of the following steps:
\begin{itemize}	
\item Generate the Kato terms provided by the perturbation 
	series~(\ref{eq:Kato_energy}) in $\nu$th order. This step is 
	independent of the particular problem under study: Once the Kato terms 
	are known, they can be used for all kinds of perturbative calculations.
	Group these terms together as far as possible, taking into account that
	odd orders never contribute, as there are no closed loops with an odd 
	number of tunneling processes.
\item Create all paths representing a closed loop with $\nu$ tunneling 
	processes. Subsume topologically identical paths to diagrams, and 
	append the correct weight factors.
\item For each diagram, go through all permutations of the individual 
	processes; for each particular sequence thus obtained, determine 
	those Kato terms which match it. Compute the corresponding matrix 
	elements, including the respective energy denominators. Sum up the 
	contributions of all sequences and all diagrams.
\end{itemize}
In high orders this procedure becomes more and more cumbersome, as the 
order~$\nu$ enters factorially, and both the number of diagrams and 
the number of Kato terms grows roughly exponentially with~$\nu$. 
Table~\ref{tab:E_dias_kato} demonstrates this for dimensionalities $d=1$, 
$2$, and $3$. Nonetheless, a considerable advantage offered by this scheme 
consists in the fact that its implementation on a computer is straightforward, 
while a representation of the entire Hilbert space is not necessary. The 
strategy of determining the contributions to the perturbation series from 
all possible paths on the lattice then allows us to treat filling factors 
and dimensionalities which are difficult to reach by other approaches. 
In order to clarify the above steps, we give an explicit example.
\begin{table}
\caption{Number of diagrams required by the energy correction for lattice 
	dimensionality~$d$, and number of Kato terms, vs.\ the order~$\nu$. If 
	the perturbation vanishes to first order, one is left with the reduced 
	number of Kato terms. This number is further diminished in case of the 
	energy correction (last column), because an even number of tunneling 
	processes has to appear between two projection operators~$S^0$.
	\label{tab:E_dias_kato}}
\begin{ruledtabular}
	\begin{tabular}{r|r|r|r|r|r|r}		
	$\nu $ & \multicolumn{3}{r|}{No. of diagrams} &
	\multicolumn{3}{r}{No. of Kato terms} \\ 
	\hline
	& $d=1$ & $d=2$ & $d=3$ & general & reduced & energy relevant \\
	\hline 
	 2 &  1 &   1 &   1 &     1 &    1 &    1 \\
	 4 &  2 &   3 &   3 &     4 &    2 &    2 \\
	 6 &  3 &   7 &   7 &    22 &    7 &    6 \\
	 8 &  6 &  29 &  29 &   119 &   26 &   18 \\
	10 & 10 & 121 & 127 &   627 &   97 &   57 \\ 
	12 & 20 & 698 &     &  3216 &  357 &  175 \\ 
	14 & 36 &     &     & 16169 & 1297 &  546 \\
	16 & 72 &     &     & 79876 & 4628 & 1672 \\
	\end{tabular}
\end{ruledtabular}
\end{table}

\subsection{Example} 
\label{subsec:Example}

\begin{figure} 
\includegraphics[scale=0.5]{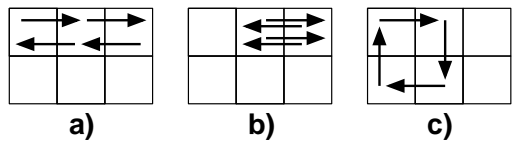}
\caption{Diagrams contributing in fourth order to the energy correction. 
 	We focus on diagram~a) in our example calculation. For lattice
	dimensionality~$d$, the respective weight factors are 
	a)~$3d(2d-1)/3$, b)~$2d/2 $, and c)~$2d(2d-2)/4 $.}
\label{fig:diagrams_E_4th_order}
\end{figure}

Let us determine the fourth-order correction of the ground-state energy due to 
the perturbation given by $H_{\rm tun}$. As this perturbation~(\ref{eq:V_tun}) 
is not diagonal in the Fock basis, the first and the second term in 
Eq.~(\ref{eq:E_4th}) vanish. The remaining Kato terms are
\begin{eqnarray}
	E_{\mathbf{m}}^{(4)} = &   &
	\langle \mathbf{m}|V S^2 V S^0 V S^1 V |\mathbf{m} \rangle 
	\nonumber\\
      + & &\langle \mathbf{m}|V S^1 V S^1 V S^1 V |\mathbf{m} \rangle \; .
\label{eq:E_4th_2}	
\end{eqnarray} 
In general, if the perturbation does not contribute to first order, the number 
of Kato terms can be significantly reduced, which leads to a substantial 
computational speedup. For calculating the energy correction one can actually 
reduce the number of terms still further, since only an even number of 
tunneling processes can appear between two projection operators~$S^0$. As 
shown in Tab.~\ref{tab:E_dias_kato}, this approximately halves the number of 
terms required in tenth order. In our example, we have to evaluate the three 
diagrams depicted in Fig.~\ref{fig:diagrams_E_4th_order}, and to determine 
their weight factors. 
\begin{figure} 
\includegraphics[scale=0.65]{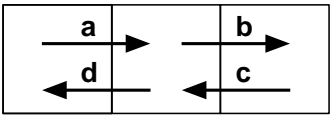}
\caption{Fourth-order diagram with individual tunneling processes labeled 
	$a$, $b$, $c$, and $d$, as considered in the example calculation.}
\label{fig:diagram_E_4th_example}
\end{figure}

We restrict ourselves to the calculation of the diagram~a) listed in  
Fig.~\ref{fig:diagrams_E_4th_order}, and denote the individual tunneling 
processes by $a$, $b$, $c$, and $d$, as indicated in 
Fig.~\ref{fig:diagram_E_4th_example}. For this computation a system with 
three lattice sites suffices. The ground state then is represented by 
$|\mathbf{m} \rangle = |g,g,g \rangle$, with filling factor~$g$. Out of a 
total number of $4!=24$ permutations, the first sequence to be treated here 
is $(a,b,c,d)$, leading to the following succession of intermediate states:
\begin{eqnarray}
	|g,g,g \rangle  
	& \stackrel{a}{\curvearrowright} & |g-1,g+1,g \rangle \nonumber\\
	& \stackrel{b}{\curvearrowright} & |g-1,g,g+1 \rangle \nonumber\\
	& \stackrel{c}{\curvearrowright} & |g-1,g+1,g \rangle 
	\stackrel{d}{\curvearrowright} 
	|g,g,g \rangle \; .
\end{eqnarray} 
Invoking the familiar ladder relations
\begin{eqnarray}
	\hat{a}^{\phantom \dagger} |n\rangle & = & \sqrt{n} |n-1\rangle 
	\nonumber\\
	\hat{a}^{\dagger} |n\rangle &=& \sqrt{n+1} |n+1\rangle  
\end{eqnarray} 
for bosonic annihilation and creation operators, the factors acquired by 
tunneling combine to $g(g+1)^3(J/U)^4$. Because the initial state does not 
occur as an intermediate state here, this particular sequence does not 
match the first term in Eq.~(\ref{eq:E_4th_2}), so that only the second one 
contributes. As one particle-hole pair is present in each intermediate state, 
the three individual energy denominators are 
$E_{\mathbf{m}}^{(0)} - E_i^{(0)} = -1$ 
(in multiples of the pair-interaction energy $U$). The full energy denominator 
therefore is given by $(-1)(-1)(-1) = -1$. Thus, the contribution to the 
energy correction provided by the sequence $(a,b,c,d)$ is
\begin{equation}
	\Delta E_{(a,b,c,d)} = -g(g+1)^3 \left(\frac{J}{U}\right)^4  \; .
\end{equation} 
Next, we treat the sequence $(a,d,b,c)$:
\begin{eqnarray}
	|g,g,g \rangle 
	& \stackrel{a}{\curvearrowright} & |g-1,g+1,g \rangle \nonumber\\
	& \stackrel{d}{\curvearrowright} & |g,g,g \rangle     \nonumber\\
	& \stackrel{b}{\curvearrowright} & |g,g-1,g+1 \rangle
	\stackrel{c}{\curvearrowright} |g,g,g \rangle \; .
\end{eqnarray} 
Here the initial state is recovered after the second tunneling process, 
leading to a contribution of the first term of Eq.~(\ref{eq:E_4th_2}), 
whereas the second one does not match. The prefactor due to tunneling now 
is $g^2(g+1)^2 (J/U)^4$; the energy denominator becomes $(-1)^2(-1) = -1$. 
Since $S^0$ yields another factor of $-1$ (see Eq.~(\ref{eq:Kato_S})), the 
contribution of this sequence reads as
\begin{equation}
	\Delta E_{(a,d,b,c)} = g^2(g+1)^2 \left(\frac{J}{U}\right)^4 \; .
\end{equation} 
In the same manner, the other 22~permutations of the processes $a$, $b$, $c$, 
and $d$ have to be evaluated. Summing up all the resulting contributions, 
multiplying by the weight factor $d(2d-1)$ pertaining to this particular
diagram, and then adding the other two diagrams with their respective weight 
factors, one arrives at the total energy correction in fourth order.

\subsection{Ground-state expectation values} 
\label{subsec:GS_exp}

The technique introduced above also allows one to calculate expectation 
values $\langle H_1 \rangle$ of observables $H_1$ in the ground state of 
the homogeneous Bose-Hubbard model as expansions in the tunneling 
strength~$J/U$. Considering an extended Hamiltonian
\begin{equation}
	H = H_0 + \lambda H_{\rm tun} + \eta H_1 \; ,
\label{eq:H_expectation}	
\end{equation} 
its ground-state energy generically possesses an expansion of the form
\begin{equation}
	E = \sum_{n,m} \lambda^n \eta^m E^{(n,m)} \; .
\end{equation} 
Denoting the ground-state wave function of the full 
system~(\ref{eq:H_expectation}) by $|\psi(\lambda, \eta) \rangle$, 
the Hellmann-Feynman theorem states
\begin{equation}
	\frac{\partial}{\partial \eta} E 
	= \langle \psi(\lambda, \eta) | \frac{\partial H}{\partial \eta} 
	| \psi(\lambda, \eta) \rangle \; ,
\end{equation} 
implying
\begin{equation}
	\sum_{n,m} m \, \lambda^n \eta^{m-1} E^{(n,m)} = 
	\langle \psi(\lambda,\eta) | H_1 | \psi(\lambda,\eta) \rangle
\end{equation} 
and thus resulting in
\begin{equation}
	\sum_{n}  E^{(n,1)} 
	= \langle \psi(1,0) | H_1 | \psi(1,0) \rangle
	\equiv \langle H_1 \rangle \; .
\end{equation}
This means that an implementation of Kato's perturbation series can be used
for computing the desired ground-state expectation values by considering 
the perturbation $V = H_{\rm tun} + H_1$ to first order in $H_1$: Each 
process chain appearing to order $n = \nu + 1$ in the perturbation series 
then contains $\nu$ tunneling events described by~$H_{\rm tun}$, and only 
one process~$H_1$. 
 
For further details concerning the process chain approach we refer to 
Ref.~\onlinecite{Eckardt08}.

\section{Results} 
\label{sec:results}

\subsection{Energy corrections} 
 
As the preceding example has shown, the process chain approach in principle 
works for lattices of any dimensionality, with arbitrary filling factor $g$, 
but with increasing order it quickly becomes impracticable to write down the 
resulting terms by hand, since their number proliferates rapidly, and it is 
unlikely that they combine to yield a simple expression. However, a numerical 
implementation on a computer is technically feasible and straightforward.

\begin{figure}
\includegraphics[scale=0.3,angle=-90]{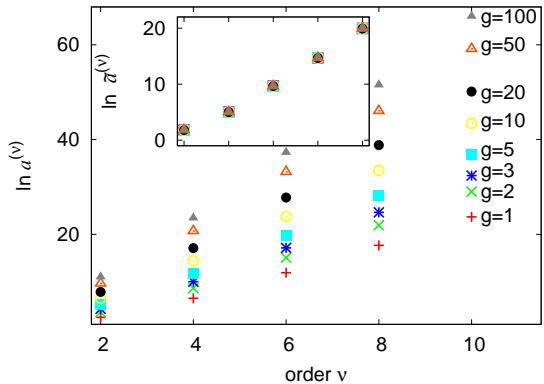}
\caption{(Color online) Logarithm of the coefficients $a^{(\nu)}$ for the 3D 
	Bose-Hubbard system with filling factors $g = 1,2,3,5,10,20,50,100$, 
	as defined in Eq.~(\ref{eq:E_series}). The inset shows the logarithm 
	of the scaled coefficients~(\ref{eq:a_scaled}), which are almost 
	independent of~$g$. The coefficients grow approximately exponentially 
	with the order~$\nu$. The results for the 2D system are qualitatively 
	similar to these.}
\label{fig:coeffs_e_d3_log}	
\end{figure}

With our current implementation we are able to calculate energy corrections 
per lattice site for the one-, two-, and three-dimensional (1D, 2D, and 3D) 
Bose-Hubbard model up to 10th order (12th order in the 1D case) in the 
tunneling coupling~$J/U$ for any integer filling factor~$g$ in the form
\begin{equation}
	\frac{E - E^{(0)}_{\mathbf{m}}}{M} = 
	-\sum_{\nu=2} a^{(\nu)}(g) \left(\frac{J}{U}\right)^{\nu} \; .
\label{eq:E_series}
\end{equation} 
Our data for the 1D system with $g = 1$ agree accurately with the results 
reported by Damski and Zakrzewski,~\cite{DamskiEtAl2006} who have performed 
a high-order symbolic perturbative expansion for this particular situation. 
In the 2D and the 3D case the coefficients $a^{(\nu)}(g)$ grow to good 
approximation exponentially with the order~$\nu$, as 
Fig.~\ref{fig:coeffs_e_d3_log} demonstrates. It is of interest to observe 
that scaling these coefficients $a^{(\nu)}(g)$ by factors $\sqrt{g(g+1)}^{\nu}$
leads to data
\begin{equation}
	\widetilde{a}^{(\nu)} = \frac{a^{(\nu)}(g)}{\sqrt{g(g+1)}^{\nu}}	
\label{eq:a_scaled}	
\end{equation} 
which are almost independent of the filling factor, as witnessed by the inset 
in Fig.~\ref{fig:coeffs_e_d3_log}. 
This is intuitively intelligible, since $\sqrt{g(g+1)}$ is a typical factor 
accompanying a tunneling process on a lattice which contains $g$~particles 
per site on the average. Therefore, one can approximately transform the 
coefficients $a^{(\nu)}(g_1)$ pertaining to one filling factor~$g_1$ 
to those referring to another factor~$g_2$:
\begin{equation}
	a^{(\nu)}(g_1) \approx 
	\left( \frac{ g_1(g_1+1)}{g_2(g_2+1) }\right)^{\nu/2}  
	a^{(\nu)}(g_2) \; .
\end{equation} 
This relation even is exact in second order, whereas small deviations occur 
in higher orders, for which it still remains a very good estimate. Naturally,
the largest deviation from this scaling behavior occurs for filling factor 
$g = 1$, as will repeatedly become visible in our data.

\begin{figure}
\includegraphics[scale=0.3,angle=-90]{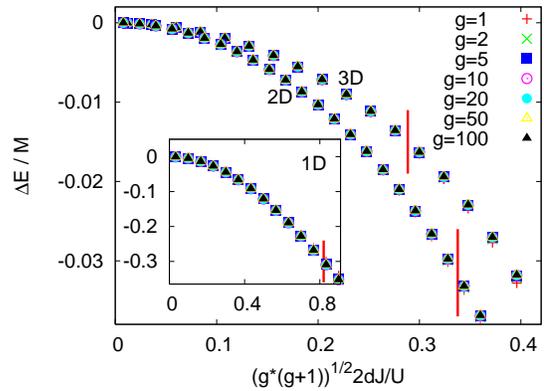}
\caption{(Color online) Ground-state energy correction per site 
	$(E - E_{\mathbf{m}}^{(0)})/M$ in multiples of the pair interaction 
	energy~$U$ for filling factors $g = 1,2,5,10,20,50,100$  for the 
	3D (upper data points),  2D (lower data points), and the 1D system 
	(inset). Vertical lines indicate the respective critical value 
	$(J/U)_{\rm c}$ for the Mott insulator-to-superfluid transition 
	with $g = 1$ (see Sec.~\ref{sec:phase}). As a result of the 
	scaling~(\ref{eq:E_series_g}), data points for different~$g$ fall 
	almost on top of each other.}
\label{fig:E_d123}	
\end{figure}

With this observation in mind, we write 
\begin{equation}
	\frac{E - E_{\mathbf{m}}^{(0)}}{M} = 
	-\sum_{\nu=2} \widetilde{a}^{(\nu)} 
	\left(\sqrt{g(g+1)}\frac{J}{U}\right)^{\nu} \; .
\label{eq:E_series_g}	
\end{equation} 
Hence, when plotting in Fig.~\ref{fig:E_d123} the energy corrections as 
functions of $\sqrt{g(g+1)} J/U$, graphs originating from different filling 
factors~$g$ practically lie on top of each other. Of course, when keeping 
$\sqrt{g(g+1)} J/U$ constant while increasing $g$, the zeroth-order term 
$E_{\mathbf{m}}^{(0)}/M = g(g-1)/2 - \mu/U$ becomes dominant, and the 
corrections become relatively small.

\subsection{Atom-atom correlation function} 

\begin{figure}
\includegraphics[scale=0.5]{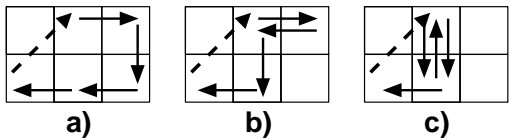}
\caption{Diagrams of fourth order in $J/U$ required for calculating 
 	$C(\vec{r})$ with $\vec{r} = [1,1,0]$. The associated weight factors 
	are a)~$12d-20$, b)~$12d-8$, and c)~$4$. The dashed arrow, pointing 
	from site~$j$ to site~$i$, describes the action of the 
	operator~(\ref{eq:H1atat}), while the solid arrows again correspond
	to nearest-neighbor tunneling processes~(\ref{eq:Htun}).}
\label{fig:diagram_C11_4th}
\end{figure}

The atom-atom correlation function is defined by 
\begin{equation}
	C(\vec{r}_{i,j}) \equiv C_{i,j} =
	\langle \hat{a}_i^{\dagger} \hat{a}_{j}^{\phantom \dagger} \rangle
\label{eq:C_ij}
\end{equation} 
with a lattice vector $\vec{r}_{i,j}$ pointing from site $j$ to site $i$. 
Since the system is invariant under translations by integer multiples of 
lattice vectors, $C_{i,j}$ depends only on $\vec{r}_{i,j}$, but not on the 
individual sites $i$ and~$j$. In the limit $J/U \to 0$ the correlation 
function $C_{i,j}$ vanishes for any non-zero $\vec{r}_{i,j}$, whereas one 
has $C_{i,j} = g$ in the BEC limit $J/U \to \infty $, when all particles 
condense into the lowest Bloch state.~\cite{DamskiEtAl2006} The calculation 
of the expectation values~(\ref{eq:C_ij}) provides an example of the 
strategy outlined in Subsec.~\ref{subsec:GS_exp}, setting
\begin{equation}
	H_1 = \hat{a}_i^{\dagger} \hat{a}_{j}^{\phantom \dagger} \; . 
\label{eq:H1atat}
\end{equation}	
Hence, we can employ the same implementation of the perturbation series as 
used for the energy correction; only the diagrams have to be adapted. Each 
process chain now has to contain one process~(\ref{eq:H1atat}), which has 
to be strictly distinguished from the nearest-neighbor tunneling processes 
described by $H_{\rm tun}$ even if $i$ and $j$ label adjacent sites. 
We therefore depict this process $H_1$ by a dashed arrow. 
In Fig.~\ref{fig:diagram_C11_4th} we display the diagrams contributing in 
fourth order of $J/U$ to $C([1,1,0])$. Because one operator 
$H_1 = \hat{a}_i^{\dagger} \hat{a}_{j}^{\phantom \dagger}$ appears 
in each process chain, the required order of perturbation theory is 
$n = \nu + 1$, where $\nu$ signals the number of ordinary tunneling processes 
$H_{\rm tun}$, as before. When determining the weight factor of a given 
diagram of this kind, no division by the number of sites occurs, because 
$i$ and $j$ distinguish specific sites.
 
\begin{figure}
\includegraphics[scale=0.3,angle=-90]{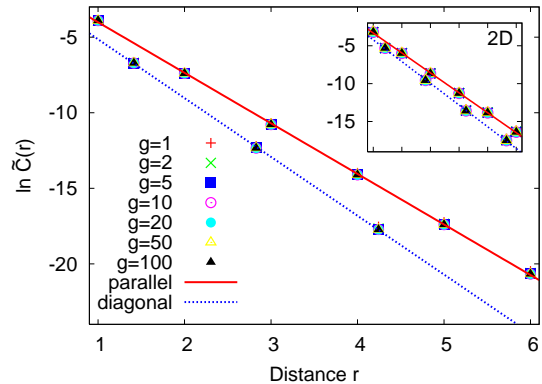}
\caption{(Color online) Logarithm of the scaled atom-atom correlation function 
	$\widetilde{C}(\vec{r}) = C(\vec{r})/\sqrt{g(g+1)}$ for $d = 3$ at 
	$J/U = 0.01/\sqrt{g(g+1)}$, with various filling factors~$g$. The 
	corresponding data for $d = 2$ and $J/U = 0.02/\sqrt{g(g+1)}$ are 
	shown in the inset. Due to the scaling, data points for different~$g$ 
	lie almost on top of each other. The decay of the correlations is 
	quite well described by exponential functions, as testified by the 
	linear fits. Correlations along the diagonal (dotted lines) decay 
	quicker than those parallel to a main axis (full lines).}
\label{fig:C_r_d23}	
\end{figure}

\begin{figure}
\includegraphics[scale=0.3,angle=-90]{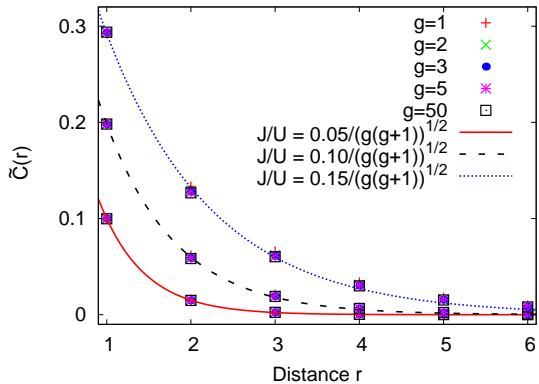}
\caption{(Color online) Scaled atom-atom correlation 
	$\widetilde{C}(\vec{r}) = C(\vec{r})/\sqrt{g(g+1)}$ 
	for the 1D system with $g=1,2,3,5,50$ at $J/U = 0.05/\sqrt{g(g+1)}$ 
	(full line), $J/U = 0.10/\sqrt{g(g+1)}$ (dashed line), and 
	$J/U = 0.15/\sqrt{g(g+1)}$ (dotted line). The lines are 
	exponential fits of the form $\beta \exp{\left(-\alpha r \right)}$, 
	with parameters $\alpha$ and $\beta$ determined for $g = 4$. 
	With increasing tunneling parameter $J/U$ the quality of the fit 
	becomes less good.}
\label{fig:C_r_d1}	
\end{figure}

We have concentrated our investigations on correlations along a line parallel 
to a principal axis of the lattice, and along a diagonal in a main lattice 
plane, as corresponding to lattice vectors
\begin{eqnarray}
	 \vec{r}_{i,j} & = 
	 \left[\begin{matrix}
	 	s, & 0, & 0
	 \end{matrix} \right] \; , & s = 1, 2, \ldots, 6 \; , 
	 \nonumber \\
	 \vec{r}_{i,j} & =
	 \left[\begin{matrix}
	 	t, & t, & 0
	 \end{matrix} \right] \; , & t = 1, 2, 3 \; ,
\label{eq:lattice_vectors}
\end{eqnarray} 
where $s$ and $t$ are given as multiples of the lattice constant. Depending 
on $\vec{r}_{i,j}$, only even or only odd orders contribute. Our current 
implementation is able to handle the expansion up to $10$th order in $J/U$ 
for dimensionalities $d = 3$ and $d = 2$, and up to $11$th order for $d = 1$. 
For the 3D system the number of the diagrams encountered is stated in 
Appendix~\ref{app:number_of_dias}. As in the case of the energy correction, 
the coefficients grow approximately exponentially with the number $\nu$ 
of ordinary tunneling processes. Our findings for the 1D system with 
unit filling ($g = 1$) again perfectly match the expansion reported in 
Ref.~\onlinecite{DamskiEtAl2006}. Moreover, once again the data can be scaled 
such that they become almost independent of the filling factor~$g$. To this 
end, we divide the correlation function by the leading density dependence, 
and plot $\widetilde{C}(\vec{r}_{i,j}) = C(\vec{r}_{i,j})/\sqrt{g(g+1)}$ 
for fixed scaled tunneling parameter 
$\sqrt{g(g+1)}J/U$ vs.\ $r_{i,j} = |\vec{r}_{i,j}|$, employing the Euclidean 
norm. At least for sufficiently small $\sqrt{g(g+1)} J/U$, we then find a 
beautiful exponential decay of the correlations with distance, depending on 
the direction considered: 
\begin{equation}
	\widetilde{C}(\vec{r}_{i,j}) \propto 
	\exp\!\big(-\alpha(J/U) r_{i,j}\big) \; .
\end{equation} 
In Fig.~\ref{fig:C_r_d23} we display logarithms of such scaled atom-atom 
correlations $\widetilde{C}(\vec{r}_{i,j})$ for $d = 3$ and  $d = 2$, 
together with linear fits. The correlations along the lattice axis are 
slightly larger than those along the diagonal. 
Figure~\ref{fig:C_r_d1} depicts $\widetilde{C}(\vec{r}_{i,j})$ for the 
1D case, for three scaled tunneling parameters. As expected, lower tunneling 
rates lead to quicker decays of the correlations. Again the scaling works 
remarkably well here, mapping the data for different filling factors nearly 
onto each other. Such an exponential decay of 1D correlations in the regime 
of low tunneling rates has also been observed with DMRG methods for distances 
up to 20 lattice constants by Kollath~\emph{et al.}~\cite{KollathEtAl04}

\begin{figure}
\includegraphics[scale=0.3,angle=-90]{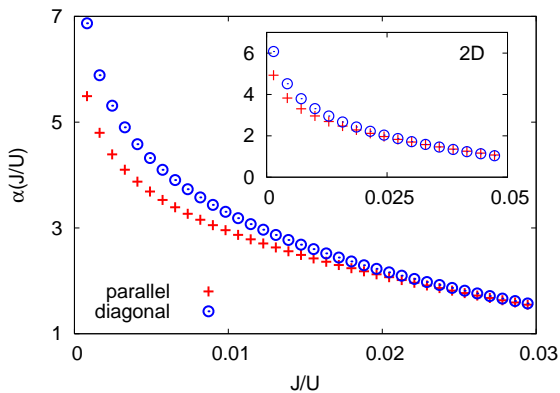}
\caption{(Color online) Correlation decay constants $\alpha(J/U)$ for $d = 3$ 
	vs.\  tunneling coupling~$J/U$, as determined tentatively from fits 
	to the correlation functions. Observe that the coefficients for 
	different directions (parallel and diagonal) converge with 
	increasing~$(J/U)$. The inset shows the corresponding results for 
	$d = 2$. All data have been computed for unit filling ($g = 1$).}
\label{fig:alpha_d23}	
\end{figure}

Figure~\ref{fig:alpha_d23} shows the evolution of the two decay constants 
$\alpha$ (parallel and diagonal) with increasing $J/U$ for the 3D and the
2D system with unit filling, determined from the slopes of linear fits to 
logarithmic plots similar to Fig.~\ref{fig:C_r_d23}. For large~$J/U$ the 
data should be considered as tentative only, since the quality of the fit
deteriorates then. While the perturbative expansion cannot be expected to 
be valid beyond the critical hopping strength which marks the transition 
to a superfluid (with $g = 1$, one finds $(J/U)_{\rm c} \approx 0.034$
for the 3D system and $(J/U)_{\rm c} \approx 0.059$ for the 2D case, see  
Refs.~\onlinecite{ElstnerMonien99,CapogrossoSansone07,CapogrossoSansone08} 
and Sec.~\ref{sec:phase}), and the true decay constants are supposed to 
vanish at that point, it is interesting to observe in Fig.~\ref{fig:alpha_d23}
that the tentative data obtained for the two directions converge with 
increasing $J/U$.

\subsection{Density-density correlations} 

\begin{figure}
\includegraphics[scale=0.5]{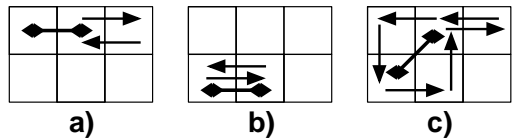}
\caption{Diagrams for calculating the density-density correlation 
	$D(\vec{r})$. Subfigures~a) and b) depict the second-order 
	diagrams required for $D([1,0,0])$, with associated weight factors 
	a)~$2(2d-1)$ and b)~$1$. The linked diamonds denote the operator 
	$\hat{n}_i \hat{n}_j$, while arrows represent nearest-neighbor 
	tunneling. Subfigure~c) depicts a more complicated diagram for 
	$D([1,1,0])$ of 6th order in $J/U$.}
\label{fig:diagram_D1_2th}
\end{figure}

Similar to the atom-atom correlation $C(\vec{r}_{i,j})$, we investigate the 
density-density correlation 
\begin{equation}
  	D(\vec{r}_{i,j}) \equiv D_{i,j} 
	= \langle \hat{n}_i \hat{n}_j \rangle \; .
\end{equation} 
Besides the nearest-neighbor tunneling processes $H_{\rm tun}$, every 
process chain now contains one operator $H_1 = \hat{n}_i \hat{n}_j$, 
sketched in the diagrams by two diamonds connected by a line, as illustrated 
in Fig.~\ref{fig:diagram_D1_2th}. 
Because this process $H_1$ does not change the particle number and 
leaves the state it acts on unaltered, the diagrams for $D_{i,j}$ always 
contain an even number of ordinary tunneling processes, as necessary for 
generating closed loops, and therefore only even orders in $J/U$ contribute. 
Since $\langle \mathbf{m}| H_1| \mathbf{m} \rangle$ does not vanish, we 
cannot reduce the number of Kato terms as much as was possible in the case of 
the energy correction and the atom-atom correlation, leaving us with higher  
computational effort. Moreover, the number of diagrams is larger than in 
the previous situations, as shown in Appendix~\ref{app:number_of_dias} for 
$d = 3$. We compute density-density correlations up to order eight in the 
tunneling parameter~$J/U$ for the 2D and the 3D system, and up to order ten 
in the 1D case.

\begin{figure}
\includegraphics[scale=0.3,angle=-90]{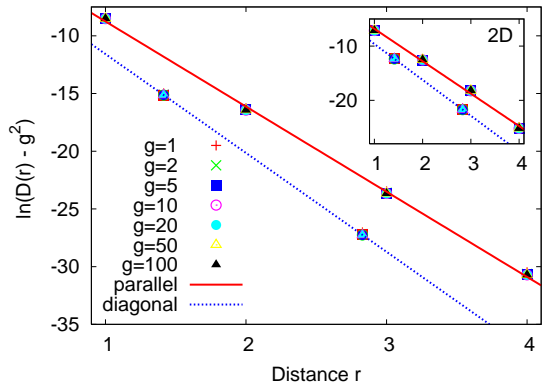}
\caption{(Color online) Logarithm of the correction 
	$\Delta D(\vec{r}) = D(\vec{r}) - g^2$ to the zeroth-order 
	density-density correlation for $d = 3$ at $J/U = 0.01/\sqrt{g(g+1)}$. 
	Because of this scaling, data points for different filling factors~$g$ 
	lie almost on top of each other. The decay of these corrections is 
	quite well described by exponential functions, as emphasized by the 
	linear fits. The inset shows data for $d = 2$ with 
	$J/U = 0.02/\sqrt{g(g+1)}$.}
\label{fig:D_r_d23}	
\end{figure}

For $d = 3$ the corrections $\Delta D(\vec{r}_{i,j}) = D(\vec{r}_{i,j}) - g^2$ 
to the zeroth-order value $g^2$ obtained for $J/U$ = 0 are fairly small, as 
exemplified in Fig.~\ref{fig:D_r_d23}. In the BEC-limit ($J/U \to \infty$), 
the density-density correlations again are given by $D_{i,j} = g^2$, assuming
large systems ($M \to \infty$). For small tunneling parameter~$J/U$ we observe 
an exponential decay of $\Delta D(\vec{r}_{ij})$ with increasing distance 
$r_{i,j}$, as in Fig.~\ref{fig:D_r_d23}. Similar to the case of the atom-atom 
correlations, the decay constants depend on the direction: ``Diagonal'' 
correlations tend to decay faster with distance than ``parallel'' ones.

\subsection{Occupation number fluctuations} 

\begin{figure}
\includegraphics[scale=0.47]{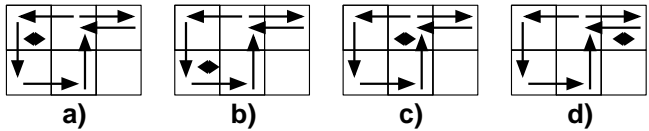}
\caption{Set of diagrams of sixth order in $J/U$ for calculating 
	$\langle \hat{n}_i^2 \rangle$. The operator $\hat{n}_i^2 $ is 
	marked by a diamond, which is added here in four topologically 
	different ways to a sixth-order diagram for the energy correction. 
	This leads to four different diagrams contributing to the 
	perturbation series in seventh order.}
\label{fig:diagram_n_6th}
\end{figure}

The squared fluctuations of the site-occupation numbers are given by the
variance
\begin{equation}
	\left( \Delta \hat{n} \right)^2 
	=  \langle \hat{n}^2 \rangle - \langle \hat{n} \rangle^2 \; .
\label{eq:n_fluctutation}
\end{equation} 
Due to the homogeneity of the Bose-Hubbard system, this quantity is 
independent of the site index, and the expectation value of the number 
operator $\hat{n}$ is just the filling factor~$g$. Thus, for determining 
the variance~(\ref{eq:n_fluctutation}) we need to know 
$\langle \hat{n}^2 \rangle$, and therefore generate our diagrams such 
that besides ordinary tunneling processes $H_{\rm tun}$ one operator 
$H_1 = \hat{n}_i^2$ appears. Because this process does not alter 
the system's state, again all Kato terms have to be evaluated, as in the 
case of the density-density correlation. Although the diagrams now look very 
similar to the ones encoding the energy correction, their number is much
higher when considering equal numbers~$\nu$ of tunneling processes, as
revealed by Tab.~\ref{tab:number_of_dias} in Appendix~\ref{app:number_of_dias}.
The reason for this increase is evident in Fig.~\ref{fig:diagram_n_6th}: 
The topology of the diagrams becomes more complex by introducing the 
additional operator~$\hat{n}_i^2$, depicted by a diamond at site~$i$. In 
the example shown in Fig.~\ref{fig:diagram_n_6th}, one diagram contributing 
in sixth order of the tunneling parameter~$J/U$ to the energy correction 
gives rise to four different diagrams for the calculation of 
$\langle \hat{n}^2 \rangle$.

\begin{figure}
\includegraphics[scale=0.3,angle=-90]{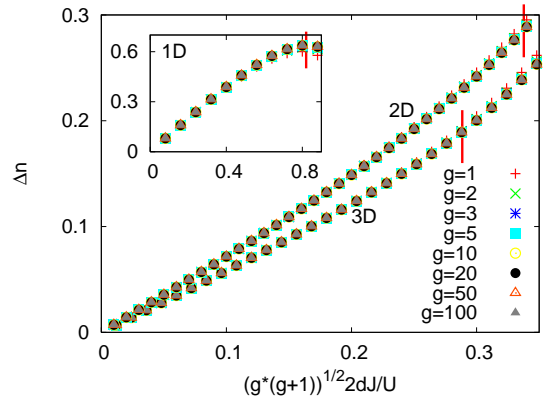} 
\caption{(Color online) Occupation number fluctuations $\Delta \hat{n} $ for 
	filling factors $g = 1$, $2$, $3$, $5$, $10$, $20$, $50$, and $100$ 
	for the 3D (lower data points), the 2D (upper data points), and the 
	1D system (inset). Vertical lines mark the critical hopping strength 
	$(J/U)_{\rm c}$ for the Mott insulator-to-superfluid transition with
	unit filling. Plotted vs.\ the scaled parameter $2d\sqrt{g(g+1)}J/U$, 
	data points for different filling factors fall onto each other. For 
	low hopping strength, the fluctuations grow linearly with $J/U$.} 
\label{fig:N_Fluctuations_d123}	
\end{figure}

We were able to determine the expansion for $\langle \hat{n}^2 \rangle$ 
up to order ten in the tunneling parameter $J/U$ for dimensionalities 
$d=1$, $2$, $3$. The number fluctuation $\Delta \hat{n}$ grows approximately 
linearly with $J/U$ for $J/U < (J/U)_{\rm c}$, and our scaling for different 
filling factors once more works very well in this parameter regime, as 
demonstrated in Fig.~\ref{fig:N_Fluctuations_d123}. Since the critical value 
$(J/U)_{\rm c}$ for the Mott insulator-to-superfluid transition is roughly 
proportional to $1/g$, this implies that the relative fluctuation 
$\Delta \hat{n}/\langle \hat{n} \rangle$ behaves like $1/g$ close to the
transition point. At $(J/U)_{\rm c}$ we find $\Delta \hat {n} \approx 1/2d$. 
In the superfluid regime, where the expansion in $J/U$ is no longer valid, 
the fluctuation $\Delta \hat{n}$ eventually approaches the value~$\sqrt{g}$. 

\section{The Mott-Superfluid phase transition} 
\label{sec:phase}

A further fruitful application of the diagrammatic many-body perturbation 
theory based on Kato's series~(\ref{eq:Kato_energy}) consists in the accurate 
determination of the boundary between the Mott phase and the superfluid phase 
for the homogeneous Bose-Hubbard model.~\cite{TeichmannEtAl2009} Qualitatively,
the phase diagram of the Bose-Hubbard model has been understood since the late 
eighties.~\cite{FisherEtAl89} More quantitatively, it has been intensely 
studied, e.g., by means of the strong-coupling expansion conducted by 
Freericks, Elstner, and Monien~\cite{FreericksMonien96,ElstnerMonien99} in 
one, two, and three dimensions. Recently, the quantum Monte Carlo analysis by 
Capogrosso-Sansone~\emph{et al.}~\cite{CapogrossoSansone07,CapogrossoSansone08}
has provided quasi-exact values for $g = 1$. In one dimension, fairly large 
systems even including a confining trap potential can be treated with DMRG 
techniques.~\cite{KuehnerEtAl1998,RapschEtAl1999,KollathEtAl04,KollathEtAl07} 
But so far, especially for dimensionalities $d > 1$ it has remained hard to 
obtain precise results for filling factors well above $g = 1$. Our approach 
is able to fill this gap.

\subsection{Method of effective potential} 
\label{subsec:mfl}

For locating the parameters $(J/U)_{\rm c}$ marking the quantum phase 
transition, we employ the method of the effective 
potential,~\cite{NegeleOrland98} in the formulation recently given by 
dos Santos and Pelster.~\cite{SantosPelster08} To begin with, one adds
spatially constant source and drain terms to the Bose-Hubbard 
Hamiltonian~(\ref{eq:Hamiltonian}), such that particles are created and 
annihilated with uniform strengths~$\eta$ and $\eta^*$ at each site:
\begin{equation}
	\widetilde{H}_{\rm BH}(\eta, \eta^*) =
	H_0 + H_{\rm tun} 
	+ \sum_i \left(\eta^* \hat{a}_i^{\phantom \dagger} 
	+ \eta \hat{a}_i^{\dagger}\right) \;.
\label{eq:BH_source}
\end{equation} 
Since our considerations apply for any fixed value of the chemical potential,
we do not explicitly indicate the dependence on $\mu/U$ in the following.
We now define the grand canonical free energy at zero temperature
\begin{equation}
	F(J/U, \eta,\eta^*) = \langle \widetilde{H}_{\rm BH} \rangle_{\eta}
\end{equation} 
as the ground-state expectation value of the full 
Hamiltonian~(\ref{eq:BH_source}) for finite source strength, and expand this 
expression into a power series in $\eta$ and $\eta^*$:
\begin{equation}
	F(J/U,\eta, \eta^*) = 
	M \left(f_0(J/U) + \sum_{n=1}^{\infty} c_{2n}(J/U) |\eta|^{2n} \right)
	\;.
\label{eq:free_energy}
\end{equation}
The appearance of only powers of $|\eta|^2$ reflects the fact that the free 
energy does not depend on the phases of $\eta$ and $\eta^*$. The intensive 
quantity $f_0$ denotes the free energy per lattice site in the absence of 
the sources. The coefficients appearing in the above expansion, in their turn, 
are expanded into power series in the hopping strength $J/U$, giving
\begin{equation}
	c_{2n}(J/U) = \sum_{\nu=0}^{\infty} \alpha_{2n}^{(\nu)} (J/U)^{\nu} 
	\; .
\label{eq:series_c2}
\end{equation}
Whether the system is a Mott insulator or a superfluid is determined by its
reaction to the sources. Hence, we introduce the functions
\begin{eqnarray}
	\psi(\eta,\eta^*) & = 
	\displaystyle{\frac{1}{M}\frac{\partial F}{\partial \eta^*}} &= 
	\langle \hat{a}_i^{\phantom \dagger} \rangle_{\eta} \; ,
	\nonumber\\
	\psi^*(\eta,\eta^*) & = 
	\displaystyle{\frac{1}{M}\frac{\partial F}{\partial \eta}} & = 
	\langle \hat{a}_i^{\dagger} \rangle_{\eta} \; ,
\label{eq:order_parms}
\end{eqnarray} 
where the respective second equalities are consequences of the Hellmann-Feynman 
theorem. Assuming the invertibility of these functions, we then perform a
Legendre transformation from $F$ to a function $\Gamma$ depending on $J/U$, 
$\psi$, and $\psi^*$  as independent variables:
\begin{equation}
	\Gamma(J/U,\psi,\psi^*) = F/M - \psi^* \eta - \psi \eta^* \; .
\label{eq:Gamma}
\end{equation} 
From Eqs.~(\ref{eq:order_parms}) and (\ref{eq:free_energy}) one obtains
\begin{alignat}{3}
 	\psi(\eta,\eta^*)   = & c_2 \eta   & & + 2c_4|\eta|^2\eta    
	& + \mathcal{O}(\eta^5) \; , 
	\nonumber  \\ 
	\psi^*(\eta,\eta^*) = & c_2 \eta^* & & + 2c_4|\eta|^2\eta^*  
	& + \mathcal{O}(\eta^5) \; .  
\end{alignat} 
Inverting these relations, and inserting into Eq.~(\ref{eq:Gamma}), 
one arrives at an expansion of $\Gamma$ in powers of $|\psi|^2$:
\begin{equation}
	\Gamma( J/U, \psi, \psi^*) = f_0 - \frac{1}{c_2}|\psi|^2 
	+\frac{c_4}{c_2^4}|\psi|^4 + \mathcal{O}(|\psi|^6)
	\;.
\label{eq:effpot}
\end{equation} 
Because $\eta$ and $\psi^*$, as well as $\eta^*$ and $\psi$,
constitute Legendre pairs, one also has the identities
\begin{equation}
	\frac{\partial\Gamma}{\partial \psi^*} = -\eta 
	\quad \text{and} \quad
	\frac{\partial\Gamma}{\partial \psi} = -\eta^*
	\; .
\label{eq:Gamma_psi}	
\end{equation}
Now the original Bose-Hubbard system~(\ref{eq:Hamiltonian}) is recovered from
the extended system~(\ref{eq:BH_source}) by setting $\eta = \eta^* = 0$.
Hence Eq.~(\ref{eq:Gamma_psi}) implies that the system adopts that value
$\psi_0$ which renders $\Gamma$ stationary. This is akin to a mechanical 
system adopting a configuration in which its potential is stationary,
signaling the absence of external forces, and thus motivates to dub $\Gamma$ 
as an ``effective potential''.

Unless $\mu/U$ is integer, one finds $c_2 < 0$ for sufficiently small $J/U$, 
whereas $c_4 > 0$ (see Appendix~\ref{app:c4}), so that one has $\psi_0 = 0$; 
this is characteristic for the Mott phase. Upon increasing $J/U$, the order 
parameter $\psi_0$ takes on a non-zero value when the system enters the 
superfluid phase, indicating long-range phase coherence. Hence, for any 
given value of the chemical potential the phase boundary $(J/U)_{\rm pb}$ 
is determined by that $J/U$ for which the minimum of the 
expression~(\ref{eq:effpot}) starts to deviate from $|\psi_0|^2 = 0$. 
Evidently, this occurs when the coefficient $-1/c_2$ of $|\psi|^2$ 
vanishes. We point out that $c_2$ can be regarded as a 
susceptibility~\cite{LandauLifschitzV} $\chi$, being the derivative of the 
function $\psi(\eta,\eta^*)$ with respect to the source $\eta$:
\begin{equation}
	\chi = 
	\left(\frac{\partial \psi}{\partial \eta}\right)_{\eta \to 0}
	= c_2 \; .
\end{equation} 
In effect, one has to identify that hopping parameter~$J/U$ for which the 
susceptibility $c_2$ diverges; this divergence marks the quantum phase 
transition.

\begin{figure}
\includegraphics[scale=0.5]{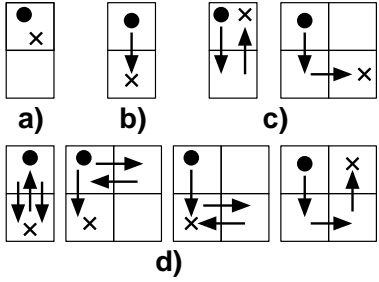} 
\caption{Diagrams for determining $c_2$ up to order~3 in the tunneling 
	parameter~$J/U$. Subfigure~a) shows the zeroth-order diagram with 
	creation ($\bullet$) and annihilation ($\times$) taking place at the 
	same lattice site, and weight factor~$1$. In b) we depict the only 
	first-order diagram; its weight factor is $2d$. The second-order 
	diagrams in subfigure~c) have weight factors $2d$ and $2d(2d-1)$. 
	The weights of the third-order diagrams in d) are (from left to right) 
	$2d$, $2d(2d-1)$, $2d(2d-1)$, and $2d(2d-1)^2$. The 
	\emph{one-way diagrams}, which acquire the largest weights for
	high dimensionality, are diagram a), b), the second diagram in~c), 
	and the last diagram in~d).}
\label{fig:diagrams_p_3}
\end{figure}

In order to compute $c_2$ by means of the process chain approach, we add 
the perturbation
\begin{equation}
	V = - J/U \sum_{\langle i,j \rangle} 
	\hat{a}_i^{\dagger} \hat{a}_j^{\phantom \dagger}
	+ \sum_i \left( \eta^* \hat{a}_i^{\phantom \dagger}  
	+ \eta \hat{a}_i^{\dagger} \right) 
\end{equation} 
to the zeroth-order Hamiltonian $H_0$, as implied by the extended 
system~(\ref{eq:BH_source}). Since $c_2$ is the coefficient of $|\eta|^2$ 
in Eq.~(\ref{eq:free_energy}), it follows by comparison of coefficients with 
Kato's series~(\ref{eq:Kato_energy}) that only chains containing one creation 
process ($\eta\hat{a}_i^{\dagger}$) and one annihilation process 
($\eta^*\hat{a}_j$) contribute to $c_2$. We adjust our diagrams by introducing 
a creation process, symbolized by a dot ($\bullet$), and an annihilation 
process, indicated by a cross ($\times$). Since the operations of creation and 
annihilation alter the particle number, the tunneling processes do not need to 
form closed loops here, in contrast to the cases examined before. This leads 
to contributions in even \emph{and\/} odd orders of $J/U$. For constructing 
the diagrams with a specified number~$\nu$ of tunneling processes, and for 
appending the correct weight factors, we generate all paths from an initial 
site to any other site which can be reached with $\nu$ nearest-neighbor 
tunneling events. The number of such paths behaves like $(2d)^{\nu}$, which 
is to a good approximation equal to the sum of all weight factors. As  
examples for the emerging diagrams, the first orders $\nu = 0$, $1$, $2$, 
and $3$ in the tunneling parameter $J/U$ are visualized in 
Fig.~\ref{fig:diagrams_p_3}.
 
Table~\ref{tab:p_d23}, which lists the number of diagrams for the 2D and 
the 3D system, shows that these numbers remain equal for both cases up to 
order $\nu=7$ (the weight factors, of course, do depend on the lattice 
dimensionality). The first difference occurs in eighth order in the 
tunneling parameter, because with eight tunneling processes it becomes 
possible to construct diagrams which connect more than four nearest neighbors 
on a hypercubic lattice. By analogy, the first difference in the number of 
diagrams between the 3D and the 4D model occurs for $\nu = 12$.
  
\begin{table}
\caption{Number of diagrams to be evaluated when calculating the phase 
	boundary for the 2D and the 3D Bose-Hubbard model to $\nu$th order 
	in the hopping parameter~$J/U$, corresponding to the order $\nu + 2$ 
	of Kato's perturbation series.}
\label{tab:p_d23} 
\begin{ruledtabular}
   \begin{tabular}{r|r|r|r|r|r|r|r|r|r|r|r}
          $\nu$ & 0 & 1 & 2 & 3 & 4  & 5  & 6  & 7   & 8   & 9   & 10   \\
   \hline $d=2$ & 1 & 1 & 2 & 4 & 10 & 22 & 58 & 140 & 390 & 988 & 2815 \\ 
   \hline $d=3$ & 1 & 1 & 2 & 4 & 10 & 22 & 58 & 140 & 394 &     &      \\ 	
   \end{tabular} 
\end{ruledtabular}
\end{table}

\subsection{Mean-field limit} 

For high dimensionality~$d$ the \emph{one-way diagrams}, which avoid 
``back and forth tunneling'' (see Fig.~\ref{fig:diagrams_p_3}), dominate 
the contributions, because their weight factors go with $(2d)^{\nu}$ to
leading power of~$d$, whereas all other diagrams contribute with lower powers.
Therefore, in the limit of large dimensionality in each order~$\nu$ \emph{only}
the one-way diagram has to be taken into account, all others possessing
negligible weight factors then. These one-way diagrams can easily be 
evaluated analytically in every given order. Since they are one-particle 
reducible, they factorize into their one-particle irreducible parts as 
follows:~\cite{SantosPelster08,Kleinert06}
\begin{eqnarray}
	\bullet \times &=& 
	(-1)^{0} \left(\bullet \times \right)^1  
\nonumber\\	
	\bullet\rightarrow \times &=& 
	(-1)^{1} \left(\bullet \times \right)^2  
\nonumber\\
	\bullet \rightarrow \rightarrow \times & = & 
	(-1)^2 \left(\bullet \times \right)^3  
\\	\vdots & & \vdots			 
\nonumber\\
	\bullet  ( \rightarrow )^\nu \times & = & 
	(-1)^\nu \left(\bullet \times \right)^{\nu+1}  
\nonumber \; .
\end{eqnarray}
Identifying the zeroth-order term $(\bullet \times)$ with $\alpha_2^{(0)}$, 
and accounting for the factor $2d$ which counts the possible directions on 
a $d$-dimensional hypercubic lattice, the representation~(\ref{eq:series_c2}) 
of the susceptibility takes the form 
\begin{equation}
  	c_2 = \alpha_2^{(0)}\sum_{\nu = 0}^{\infty} 
	\left(-2d\alpha_2^{(0)}\right)^{\nu}
	\left(\frac{J}{U}\right)^{\nu} \; .
\label{eq:geoseries_c2}	
\end{equation}
Because this series is geometric, its radius of convergence, and hence
the phase boundary, can be immediately read off from the relation  
\begin{equation}
	-2d\alpha_2^{(0)} \left(\frac{J}{U}\right)_{\rm pb} = 1 \; . 	
\end{equation}
Therefore, it only remains to compute $\alpha_2^{(0)}$ by evaluating 
the diagram $\bullet \times$. This gives rise to two permutations, which we 
write as $(\times \, , \, \bullet)$ and $(\bullet \, , \, \times)$: Either 
the creation process precedes that of annihilation, or vice versa. The only 
relevant Kato term now is $\langle \mathbf{m}| VS^1V | \mathbf{m} \rangle $; 
the respective energy denominators enforced by the linking operator $S^1$ are   
$\Delta E_{\rm particle} = E_{\mathbf{m}}^{(0)} - E_{\rm particle} = \mu/U - g$
for an extra particle, and 
$\Delta E_{\rm hole} = E_{\mathbf{m}}^{(0)} - E_{\rm hole} = -\mu/U + g-1$
for an extra hole. Thus, the two contributions figure as 
\begin{eqnarray*}
	(\times \, , \, \bullet): & &
	\sqrt{g+1} \frac{1}{\Delta E_{\rm particle}} \sqrt{g+1} 
	= \frac{g+1}{\mu/U - g} \; , \\
	(\bullet \, , \, \times): & &
	\sqrt{g} \frac{1}{\Delta E_{\rm hole}} \sqrt{g}
	= \frac{g}{-\mu/U + g-1} \; ;		
\end{eqnarray*} 
combining them yields
\begin{equation}
	\alpha_2^{(0)} = -\frac{\mu/U+1}{(\mu/U-g+1)(g-\mu/U)}
	\;.	
\end{equation} 
Putting everything together, the phase boundary in the limit of infinite 
dimensionality, as determinded from the radius of convergence of the 
series~(\ref{eq:geoseries_c2}), is given by
\begin{equation}
	2d \left(\frac{J}{U}\right)_{\rm pb} 
	= \frac{(\mu/U-g+1)(g-\mu/U)}{\mu/U+1}
	\;,	
\label{eq:mfboundary}	
\end{equation} 
which agrees exactly with the result of the mean-field calculation by 
Fisher~\emph{et al.}~\cite{FisherEtAl89} This not only clarifies why the
mean-field limit coincides with that of infinite dimensionality, but also 
gives a vivid illustration of our general approach.

\subsection{Results for lower dimensions} 

\begin{figure}
\includegraphics[scale=0.3,angle=-90]{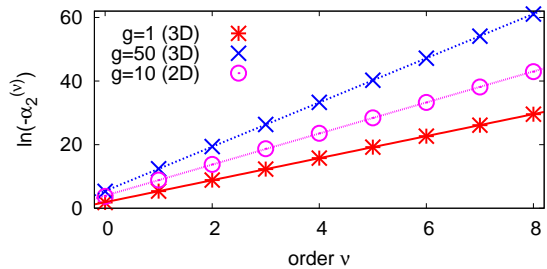}
\caption{(Color online) Logarithm of the coefficients $-\alpha_2^{(\nu)}$, 
	obtained with filling factors $g = 1$, $50$ for the 3D model, and 
	with $g = 10$ for the 2D case.}
\label{fig:coeffs_p_d23_log}	
\end{figure}

For lattice dimensionalities~$d = 2$ and $d = 3$, we have computed the 
(negative) coefficients $\alpha_2^{(\nu)}$ up to order $\nu = 8$. Again, these 
coefficients grow to good approximation exponentially with the number~$\nu$ 
of tunneling processes taken into account, as illustrated in 
Fig.~\ref{fig:coeffs_p_d23_log} for filling factors $g=1$, $50$ ($d = 3$) 
and $10$ ($d = 2$). Data for other filling factors behave similarly.

\begin{figure}
\includegraphics[scale=0.3,angle=-90]{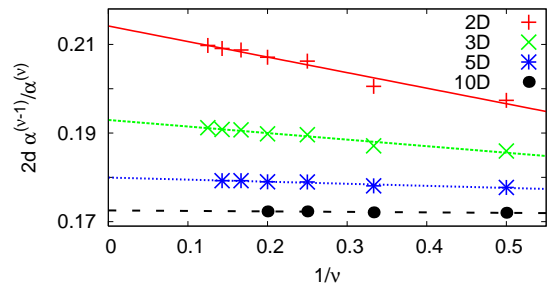} 
\caption{(Color online) Extrapolation scheme for determining the phase 
	boundary: The ratios $\alpha_2^{(\nu-1)}/\alpha_2^{(\nu)}$ are 
	plotted vs.\ $1/\nu$, and extrapolated linearly to $1/\nu=0$. Data 
	are given for dimensionalities $d = 2$, $3$, $5$, and $10$, and 
	chemical potential $\mu/U = 0.5$. Observe that the slope decreases 
	with increasing dimensionality.}
\label{fig:Extrapolation}	
\end{figure}

In the case of infinite dimensionality, the ratio 
$\alpha_2^{(\nu-1)}/\alpha_2^{(\nu)}$ stays constant, and directly yields the 
phase boundary. In contrast, for finite~$d$ this ratio still changes slightly 
with increasing $\nu$. We therefore make use of a simple extrapolation scheme, 
based on d'Alembert's ratio test:~\cite{WhittakerWatson} The  radius~$r$ of 
convergence of a power series $s=\sum_{\nu=0}^{\infty} b^{(\nu)} z^\nu$ is 
given by
\begin{equation}
	r = \lim_{\nu\to \infty}\left| \frac{b^{(\nu-1)}}{b^{(\nu)}}\right|
	\; ,
\end{equation} 
if this limit exists.
Therefore, we determine the phase boundary $(J/U)_{\rm pb}$ by plotting the 
ratios $\alpha_2^{(\nu-1)}/\alpha_2^{(\nu)}$ for orders $\nu = 1$ to $8$  
vs.\ $1/\nu$, and by extrapolating to $\nu = \infty$ by means of a linear fit. 
Figure~\ref{fig:Extrapolation} illustrates this scheme for dimensionalities
$d = 2$, $3$, $5$, and $10$, assuming unit filling. Note that the slope of 
the straight fitting lines decreases with increasing dimensionality, signaling 
the approach to the strictly geometric series present for $d = \infty$.

\begin{figure}
\includegraphics[scale=0.33,angle=-90]{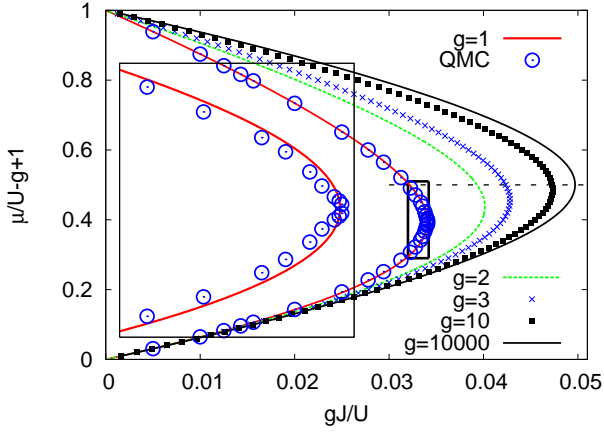}
\caption{(Color online) Mott lobes for $d = 3$, and various filling 
	factors~$g$. The dashed line marks the limit 
	$(\mu/U)_{\rm c} = g - 0.5$ attained for high~$g$. For $g = 1$, 
	QMC data~\cite{CapogrossoSansone07} are included. The inset 
	magnifies the tip of the lobe belonging to unit filling.} 
\label{fig:phase_dia}	
\end{figure}

\begin{figure}
\includegraphics[scale=0.33,angle=-90]{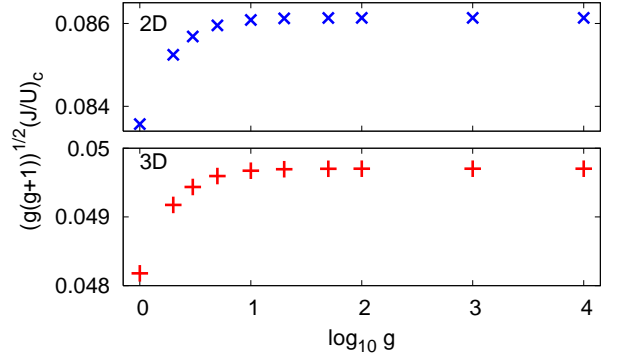}
\caption{(Color online) Scaled values $\sqrt{g(g+1)}(J/U)_{\rm c}$ of the 
	critical hopping parameter vs.\ filling factor~$g$ for $d = 2$ 
	(upper panel) and $d = 3$ (lower panel). Observe the rather fine 
	scale at the left margin.} 
\label{fig:JC_g}	
\end{figure}

This method of extrapolation also provides a reliable estimate of the 
systematic error. If we employ different selections of coefficients 
$\alpha_2^{(\nu)}$, such as those with $\nu = 2$ to $8$ or $\nu = 3$ to $8$, 
we obtain very similar values for the phase boundary, thus confirming the high 
fidelity of our data. This is shown in Tab.~\ref{tab:error_estimate}, which 
lists raw data for the critical hopping strengths $(J/U)_{\rm c}$, marking 
the position of the tip of the respective Mott lobe.
\begin{table}
\caption{Critical values of the hopping parameter $(J/U)_{\rm c}$ for
	dimensionalities $d = 2$, $3$, and $4$, and filling factor $g = 1$.
	These data were obtained from linear fits to coefficients from 
	different orders~$\nu$, as stated in the left column.}
\label{tab:error_estimate}
\begin{ruledtabular}	
\begin{tabular}{r|r|r|r}	
	$\nu $ & \multicolumn{3}{c}{$(J/U)_{\rm c}$} \\ \hline
	& $d=2$ & $d=3$ & $d=4$  \\
	\hline
	1 - 8 & 5.9093E-002 &  3.4068E-002  &  2.4131E-002\\
	2 - 8 & 5.9853E-002 &  3.4248E-002  &  2.4189E-002\\
	3 - 8 & 5.9403E-002 &  3.4092E-002  &  2.4107E-002\\
	4 - 8 & 5.9846E-002 &  3.4255E-002  &  2.4163E-002\\
	5 - 8 & 5.9482E-002 &  3.4080E-002  &  2.4093E-002
\end{tabular}
\end{ruledtabular}
\end{table}
For the 2D system we thus estimate the overall relative error to be smaller 
than 2\%, while it is smaller than 1\% for the 3D model, and reduced still 
further in the 4D case. The comparison of our results with recent data for 
$g = 1$ obtained by QMC methods~\cite{CapogrossoSansone07,CapogrossoSansone08} 
shows a remarkable agreement. Only close to the tip of the lobes small 
deviations are visible, as revealed by the inset of Fig.~\ref{fig:phase_dia}
for $d = 3$. A similar comparison for $d = 2$ can be found in 
Ref.~\onlinecite{TeichmannEtAl2009}.
The higher the filling factor, the more pronounced the model's approximate 
particle-hole symmetry~\cite{Sachdev99} becomes, which renders the Mott 
lobes symmetric with respect to the chemical potential, such that the 
critical chemical potential approaches $(\mu/U)_{\rm c} = g - 0.5$ for 
high~$g$ (see Fig.~\ref{fig:phase_dia}). Interestingly, when multiplying the 
critical hopping parameters $(J/U)_{\rm c}$ for fixed dimensionality~$d$ and
varying~$g$, as listed in Tab.~\ref{tab:J_c}, by $\sqrt{g(g+1)}$, all data 
fall within a quite narrow range, as witnessed by Fig.~\ref{fig:JC_g}.
\begin{table}
\caption{Critical values $(\mu/U)_{\rm c}$ and $(J/U)_{\rm c}$ for various 
	filling factors~$g$. For locating the tip of the respective Mott lobe, 
	$\mu/U$ has been varied in steps of $0.001$. Relative errors of 
	$(J/U)_{\rm c}$ are less than 1\% for $d = 3$, and less than 2\% for 
	$d = 2$.}
\label{tab:J_c}
\begin{ruledtabular}
\begin{tabular}{r|r|r|r|r}
      & \multicolumn{2}{c|}{$d=2$} & \multicolumn{2}{c}{$d=3$} \\ \hline
      $g$ & $(\mu/U)_{\rm c}$ & $(J/U)_{\rm c}$ & $(\mu/U)_{\rm c}$    
      & $(J/U)_{\rm c}$   \\ \hline
        1 & 0.376        & 5.909E-002   &  0.393       & 3.407E-002       \\
        2 & 1.427        & 3.480E-002   &  1.437       & 2.007E-002       \\
        3 & 2.448        & 2.473E-002   &  2.455       & 1.427E-002       \\
        4 & 3.460        & 1.920E-002   &  3.465       & 1.108E-002       \\
        5 & 4.470        & 1.569E-002   &  4.472       & 9.055E-003       \\
	6 & 5.472	 & 1.327E-002   &  5.476       & 7.657E-003	  \\
	7 & 6.476	 & 1.150E-002   &  6.479       & 6.634E-003       \\
	8 & 7.479	 & 1.014E-002   &  7.482       & 5.852E-003       \\
	9 & 8.481	 & 9.073E-003   &  8.484       & 5.235E-003       \\
       10 & 9.483        & 8.208E-003   &  9.485       & 4.736E-003       \\
       20 & 19.491       & 4.202E-003   &  19.492      & 2.425E-003       \\
       50 & 49.496       & 1.706E-003   &  49.497      & 9.842E-004       \\
      100 & 99.498       & 8.571E-004   &  99.498      & 4.946E-004       \\
     1000 & 999.500      & 8.609E-005   &  999.500     & 4.968E-005       \\
    10000 & 9999.500     & 8.613E-006   &  9999.500    & 4.970E-006
\end{tabular}
\end{ruledtabular}	
\end{table}

We point out that the calculation of the phase boundary for $d = 1$ 
requires further considerations, because of a re-entrance 
phenomenon:~\cite{KuehnerEtAl1998} For certain values of the chemical potential
the transition from the Mott insulator to the superfluid in the 1D system is 
followed by another transition back to the insulator phase upon increasing 
$J/U$, before the superfluid phase is reached again. Thus, for one value of 
$\mu/U$ there then exist three phase-bounding values of~$J/U$, which cannot 
immediately be extracted with our present procedure.

\subsection{Approaching the mean-field limit}

\begin{figure}
\includegraphics[scale=0.33,angle=-90]{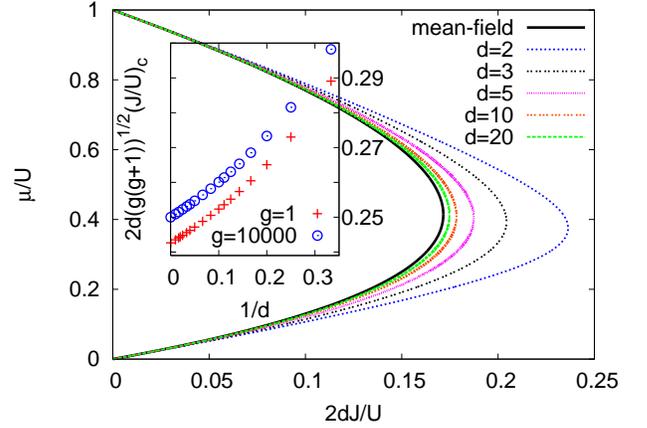}
\caption{(Color online) Phase diagram for dimensionalities $d=2$, $3$, 
 	$5$, $10$, and $20$, together with the mean-field phase boundary, 
	for $g = 1$. With increasing dimensionality the data approach the 
	mean-field prediction~(\ref{eq:mfboundary}). In the inset the scaled 
	critical hopping strength $2d \sqrt{g(g+1)} (J/U)_{\rm c}$ is plotted 
	vs.\ $1/d$ for filling factors $g = 1$ (lower data set) and 
	$g = 10000$ (upper data set).} 
\label{fig:phase_dia_d}	
\end{figure}

Although of lesser experimental relevance, it is still interesting to 
investigate systems with dimensionality $d > 3$, in order to study the 
convergence towards the mean-field limit. For higher~$d$ it becomes harder 
to obtain the necessary diagrams, and their weight factors. Nonetheless, 
we are able to treat systems of arbitrary dimensionality at least up to 
order $\nu = 4$ in the tunneling parameter $J/U$, because the corresponding 
weight factors can still be figured out by combinatorial reasoning. 
Despite this relatively low order the precision of the phase boundaries 
thus obtained is quite high for large~$d$, since the fluctuation of the 
ratio $\alpha_2^{(\nu-1)}/\alpha_2^{(\nu)}$ 
decreases significantly with increasing $d$, as illustrated by 
Fig.~\ref{fig:Extrapolation}. The resulting Mott lobes for unit filling are 
displayed in Fig.~\ref{fig:phase_dia_d} for $d=2$, $3$, $5$, $10$, and $20$, 
together with the mean-field phase boundary. In the limit $d\to \infty$ 
the critical parameter $(J/U)^{\rm mf}_{\rm c}$ can be deduced from 
Eq.~(\ref{eq:mfboundary}), giving
\begin{equation}
	2d\left(\frac{J}{U}\right)^{\rm mf}_{\rm c} = 2g+1 - 2\sqrt{g(g+1)} \; ,
\label{eq:JU_c_mf}
\end{equation} 
which scales like $1/g$ for large~$g$. Hence, the approach to the mean-field 
limit can be well monitored by plotting the phase-bounding chemical potentials
for each~$d$ vs.\ $2dJ/U$, as in the main part of Fig.~\ref{fig:phase_dia_d}. 
While the curves agree fairly well with each other at the edges of the lobes 
even for low dimensionalities, larger deviations occur around the tips. The 
convergence to the mean-field phase boundary with increasing~$d$ is clearly 
visible.
When plotting the scaled data $2d\sqrt{g(g+1)}(J/U)_{\rm c}$ for fixed~$g$
as functions of $1/d$ in order to directly highlight the approach to infinite 
dimensionality, as done in the inset of Fig.~\ref{fig:phase_dia_d} for $g=1$ 
and $g=10000$, we obtain smooth curves aiming for $1/d \to 0$ at the 
respective mean-field result determined by Eq.~(\ref{eq:JU_c_mf}). The scaled 
data belonging to different~$g$ fall into a remarkably narrow range; their 
relative separation amounts to about 3\%. In fact, for any dimensionality~$d$
and any filling factor~$g$ the critical values computed in this work are fitted by
\begin{equation}
	\left(\frac{J}{U}\right)_{\rm c} = 
	\left(\frac{J}{U}\right)^{\rm mf}_{\rm c} 
	+ \frac{0.13}{\sqrt{g(g+1)}\,d^{2.5}}
\end{equation} 
with an accuracy of about 1\%, where $(J/U)^{\rm mf}_{\rm c}$ follows from 
Eq.~(\ref{eq:JU_c_mf}). Note that even for relatively 
high~$d$ the deviation from the mean-field prediction is not negligible. 
For example, even for $d = 10$ and $g = 1$ the value of $(J/U)_{\rm c}$  
still exceeds the mean-field limit by about 4\%.

\section{Conclusion \& Outlook} 
\label{sec:conclusion} 

The essence of the high-order process chain approach is captured by the example
given in Subsec.~\ref{subsec:Example}: On the one hand, one has to generate 
the Kato terms, as dictated by the perturbation series~(\ref{eq:Kato_energy}).
This step is universal, and thus has to be done only once for all kinds of 
perturbative calculations. On the other hand, one has to construct all the 
diagrams pertaining to the particular observable under study, and their weight 
factors. These diagrams then are worked out in a Cinderella-type fashion:
Each permutation of the processes constituting a diagram has to be compared 
to the pattern of intermediate states provided by the Kato terms; the matching
permutations are evaluated, the others discarded. The bottlenecks of this
scheme are the generation of the diagrams, which poses nontrivial combinatorial
tasks in higher orders, and the factorial growth of the number of permutations
with the order of perturbation theory. While it might be feasible to optimize 
diagram-generation with the help of specifically adapted routines, the 
explosive growth of the number of permutations currently appears to limit 
straightforward numerical applications of this algorithm to about the 
twelfth order.

But still, this could open up substantial possibilities for the further 
analysis of strongly correlated quantum many-body systems. This suggestion 
is underlined not only by our calculations of the various ground-state 
correlation functions for the homogeneous Bose-Hubbard model in 
Sec.~\ref{sec:results}, but also by the comprehensive determination of its 
phase diagram in Sec.~\ref{sec:phase}. The entire set of all dimensionalities 
$d \ge 2$ and all filling factors~$g$ has been covered by a single approach, 
giving excellent agreement with previous findings in those cases for which 
accurate calculations had been performed before.    

The conceptual ease with which these results have been obtained here suggests 
that the process chain approach~\cite{Eckardt08} should also turn out useful 
for the theoretical investigation of other systems which so far are less well 
understood. We expect our strategy to work with similar success for different 
types of lattices, such as triangular or hexagonal ones, for ladder 
systems,~\cite{DonohueGiamarchi01} and for lattices with a superstructure, 
such as recently considered in Ref.~\onlinecite{BarmettlerEtAl08}.

\begin{acknowledgments} 

We thank M.~Langemeyer and V.~Steenhoff for important help with the 
generation of high-order Kato terms, and B.~Capogrosso-Sansone for 
providing the QMC data.~\cite{CapogrossoSansone07,CapogrossoSansone08}
Computer power was obtained from the GOLEM~I cluster of the Universit\"at
Oldenburg. 
N.~T.\ acknowledges a fellowship from the Studienstiftung des deutschen 
Volkes.
A.~E. thanks M.~Lewenstein for kind hospitality at ICFO-Institut de
Ci\`encies Fot\`oniques and acknowledges a Feodor Lynen research grant
from the Alexander von Humboldt foundation, as well as support by the
Spanish MEC (grant FIS2008-00784 ``TOQATA'', ESF-EUROQUAM program
FIS2007-29996-E ``FERMIX''). 
This work was further supported by the Deutsche Forschungsgemeinschaft.

\end{acknowledgments}

\begin{appendix}

\section{Number of diagrams}
\label{app:number_of_dias}
\begin{table}[h]
\caption{Number of diagrams encountered to $\nu$th order in the tunneling 
	coupling~$J/U$ when calculating the quantities considered in 
	Sec.~\ref{sec:results} for the 3D Bose-Hubbard model. An obvious
	shorthand notation is used here for the lattice vectors introduced 
	in Eq.~(\ref{eq:lattice_vectors}), such that $C(s)=C([s,0,0])$ and
	$D(t,t) = D([t,t,0])$.}
\label{tab:number_of_dias}	
\begin {ruledtabular}
\begin{tabular}{r|r|r|r|r|r|r|r|r|r|r|r}
	$\nu$       & 0 & 1 & 2 & 3 & 4 & 5 & 6  & 7 & 8 & 9 & 10 \\ \hline 
	$E$         & - & - & 1 & - & 3 & - & 7  & - & 29  & -   & 127 \\ 
	$C(1)$      & - & 1 & - & 3 & - & 10& -  & 50& -   & 281 & -   \\ 
	$C(2)$      & - & - & 1 & - & 3 & - & 15 & - & 102 & -   & 795 \\ 
	$C(3)$      & - & - & - & 1 & - & 3 & -  & 18& -   & 102 & -   \\ 
	$C(4)$      & - & - & - & - & 1 & - & 3  & - & 19  & -   & 151 \\ 
	$C(1,1)$    & - & - & 1 & - & 3 & - & 17 & - & 126 & -   & 1051\\ 
	$C(2,2)$    & - & - & - & - & 1 & - & 3  & - & 21  & -   & 190 \\ 
	$D(1)$      & 1 & - & 2 & - & 8 & - & 40 & - & 250 & -   &     \\ 
	$D(2)$      & 1 & - & 1 & - & 6 & - & 28 & - & 194 & -   &     \\ 
	$D(3)$      & 1 & - & 1 & - & 5 & - & 23 & - & 144 & -   &     \\ 
	$D(1,1)$    & 1 & - & 1 & - & 7 & - & 32 & - & 227 & -   &     \\ 
	$D(2,2)$    & 1 & - & 1 & - & 5 & - & 22 & - & 140 & -   &     \\ 
	$\hat{n}^2$ & 1 & - & 1 & - & 4 & - & 18 & - & 106 & -   & 697 \\ 
\end{tabular} 
\end {ruledtabular}
\end{table}

\section{Positivity of $c_4$}
\label{app:c4}

The method of the effective potential outlined in Subsec.~\ref{subsec:mfl}
crucially requires that the coefficient of $|\psi|^4$ in Eq.~(\ref{eq:effpot}),
and hence $c_4$, be positive. For evaluating $c_4$ within the process chain 
approach, we have to construct diagrams containing exactly two annihilation 
and two creation processes. To zeroth order in $J/U$ (fourth order of 
the perturbation series), one gets one diagram with two creations and two 
annihilations at the same site ($\bullet \bullet \times \times$), leading 
to $4!/(2!)^2=6$ permutations. Similar to Eq.~(\ref{eq:E_4th_2}), the 
corresponding Kato terms are
$\langle \mathbf{m}| V S^2 V S^0 V S^1 V | \mathbf{m} \rangle$ and  
$\langle \mathbf{m}| V S^1 V S^1 V S^1 V | \mathbf{m} \rangle$. 
The relevant energy denominators are as follows: 
\begin{alignat*}{3}
	&\text{One hole: }       & &\Delta E_{h}  & = & g-1 - \mu/U \\
	&\text{Two holes: }      & &\Delta E_{hh} & = & 2g-3 - 2\mu/U \\
	&\text{One particle: }   & &\Delta E_{p}  & = &  \mu/U -g \\
	&\text{Two particles: }  & &\Delta E_{pp} & = &  2\mu/U-(2g+1)	
	\; .
\end{alignat*} 
All these energy differences are negative, with 
\begin{eqnarray}
	\Delta E_{h} & > \Delta E_{hh} \; , \qquad 
	\Delta E_{p} & > \Delta E_{pp} \; , 
	\nonumber \\
	\Delta E_{h} & > \Delta E_{pp} \; , \qquad 
	\Delta E_{p} &> \Delta E_{hh}  \; ,
\label{eq:Denom}
\end{eqnarray}
as follows from $g-1 < \mu/U < g$. The combination of all contributions
then yields
\begin{eqnarray}
	c_4^{(0)} =& &\frac{g+1}{(\Delta E_{p})^2} 
	\left[ \frac{g+2}{\Delta E_{pp}} - \frac{g+1}{\Delta E_{p}} 
	- \frac{g}{\Delta E_{h}} \right]
	 \nonumber\\
	&+ &\frac{g}{(\Delta E_{h})^2}
	\left[ \frac{g-1}{\Delta E_{hh}} - \frac{g+1}{\Delta E_{p}} 
	- \frac{g}{\Delta E_{h}} \right] \; .
\end{eqnarray} 
Since according to the above relations~(\ref{eq:Denom}) both factors in 
square brackets are positive, the coefficient $c_4$ is manifestly positive 
to zeroth order in the tunneling parameter~$J/U$. We have investigated higher 
orders in $J/U$ numerically, and obtained only positive contributions 
$\alpha_4^{(\nu)}$.

\end{appendix}

\end{document}